\documentclass[aps,prb,twocolumn,superscriptaddress,reprint]{revtex4-2}

\usepackage{graphicx}
\usepackage{color}
\usepackage{amsmath}
\usepackage{amssymb}
\usepackage[utf8]{inputenc}
\usepackage[T1]{fontenc}
\usepackage[pdftex,colorlinks=true,linkcolor=blue,citecolor=blue,filecolor=blue,baseurl=empty,breaklinks]{hyperref}

\newcommand{\MeVfm}{\text{MeV}~\text{fm}^3}
\newcommand{\avg}[1]{\big< #1 \big>} 
\renewcommand{\vec}[1]{\ensuremath{\mathbf{#1}}} 
\newcommand{\ket}[1]{\big| #1 \big>} 
\newcommand{\bra}[1]{\big< #1 \big|} 
\newcommand{\matrixel}[3]{\big< #1 \vphantom{#2#3} \big| #2 \big| #3 \vphantom{#1#2} \big>} 
\newcommand{\lr}[1]{\left(#1\right)}

\begin{document}

\title{Matrix-product state approach to the generalized nuclear pairing Hamiltonian}

\author{Roman Rausch*}
\email[]{r.rausch@tu-braunschweig.de}
\affiliation{Technische Universit\"at Braunschweig, Institut f\"ur Mathematische Physik, Mendelssohnstra{\ss}e 3, 38106 Braunschweig, Germany}

\author{Cassian Plorin}
\affiliation{Department of Physics, University of Hamburg and The Hamburg Centre of Ultrafast Imaging, Notkestra{\ss}e 9, D-22607 Hamburg, Germany}

\author{Matthias Peschke}
\affiliation{Institute for Theoretical Physics Amsterdam and Delta Institute for Theoretical Physics, University of Amsterdam, Science Park 904, 1098 XH Amsterdam, The Netherlands}

\author{Christoph Karrasch}
\affiliation{Technische Universit\"at Braunschweig, Institut f\"ur Mathematische Physik, Mendelssohnstra{\ss}e 3, 38106 Braunschweig, Germany}

\begin{abstract}
We show that from the point of view of the generalized pairing Hamiltonian, the atomic nucleus is a system with small entanglement and can thus be described efficiently using a 1D tensor network (matrix-product state) despite the presence of long-range interactions. The ground state can be obtained using the density-matrix renormalization group (DMRG) algorithm, which is accurate up to machine precision even for large nuclei, is numerically as cheap as the widely used BCS (Bardeen-Cooper-Schrieffer) approach, and does not suffer from any mean-field artifacts.

We apply this framework to compute the even-odd mass differences of all known lead isotopes from $^{178}$Pb to $^{220}$Pb in a very large configuration space of 13 shells between the neutron magic numbers 82 and 184 (i.e., two major shells) and find good agreement with the experiment. We also treat pairing with non-zero angular momentum and determine the lowest excited states in the full configuration space of one major shell, which we demonstrate for the $N=126$, $Z\geq 82$ isotones.

To demonstrate the capabilities of the method beyond low-lying excitations, we calculate the first 100 excited states of $^{208}$Pb with singlet pairing and the two-neutron removal spectral function of $^{210}$Pb, which relates to a two-neutron pickup experiment.
\end{abstract}

\keywords{density-matrix renormalization group, DMRG, matrix-product states, MPS, nuclear pairing}

\maketitle

\section{\label{sec:introduction}Introduction}

The atomic nucleus is a many-body problem of interacting fermions and there are two main lines of approach to solve it: i) \textit{Ab initio} nuclear structure calculations are nowadays feasible up to the medium-mass and heavy regime~\cite{Hergert2020}. Since these techniques are numerically quite costly, there is a parallel interest in ii) solvable model Hamiltonians that can provide insights into nuclear structure at a much lower numerical cost.

The simplest approach at the single-particle level postulates an effective confining potential, such as the Woods-Saxon potential~\cite{Woods_Saxon_1954}. While being successful in describing various features of nuclei, it cannot account for the fact that a short-range attractive interaction prefers the formation of singlet pairs~\cite{Heyde_1994,Brink_Broglia_2005,Zelevinsky_Volya_2017}. In order to incorporate this effect, facing the complexity of the full many-body problem is unavoidable even for simplified model Hamiltonians.

A basic model to study the pairing effect was introduced by Richardson~\cite{Richardson_1966}. Its Hamiltonian reads:
\begin{equation}
\begin{split}
H = d\sum_{j=1}^L  j n_{j} -G &\sum_{jj'} a^{\dagger}_{j\uparrow} a^{\dagger}_{j\downarrow} a_{j'\downarrow} a_{j'\uparrow},
\label{eq:H_Richardson}
\end{split}
\end{equation}
where $a^{\dagger}_{j\sigma}$ is the creation operator of a spin-$1/2$ particle with spin projection $\sigma=\uparrow,\downarrow$, $n_{j}=\sum_{\sigma} a^{\dagger}_{j\sigma}a_{j\sigma}$ is the particle number operator in an orbital $j$. There are $L$ equidistant orbitals with on-site energies $\epsilon_j=jd$, where $d$ is the energy spacing. The second term with $G>0$ describes attractive pair hopping between all orbitals.

Model~\eqref{eq:H_Richardson} is integrable~\cite{Richardson_1966} and was rediscovered in the context of finite superconducting grains~\cite{vonDelft2001,Pavesic_2021}, where it can be used for quantitative predictions. In nuclear physics, it serves as a demonstration model (akin to the Ising model of condensed matter).

For a quantitative description of a nucleus, one needs to use a realistic level scheme $\epsilon_j$. Additional refinement comes from using level-resolved interaction matrix elements $G_{jj'}$:
\begin{equation}
\begin{split}
H = \sum_{j,m>0}\epsilon_{j} n_{jm} - \sum_{jj'} G_{jj'} &\sum_{m>0} \lr{-1}^{j-m} a^{\dagger}_{jm}a^{\dagger}_{j,-m} \times\\ 
&\sum_{m'>0} \lr{-1}^{j'-m'} a_{j',-m'} a_{j'm'},
\label{eq:H}
\end{split}
\end{equation}
with $n_{jm} = a^{\dagger}_{jm}a_{jm}+a^{\dagger}_{j,-m}a_{j,-m}$.
Now, $j$ is understood to be a combined index that labels the set of orbitals, each of which has a half-integer angular momentum with projection $m=-j,-j+1,\ldots,+j$ (see the Supporting Information for more details). The nucleons form singlet pairs and the factor $\lr{-1}^{j-m}$ is related to the corresponding Clebsch-Gordan coefficient $C^{00}_{jmj,-m}$. The net effect of the pairing term is that total-singlet states split off and decrease in energy, while the other angular momenta remain degenerate. The model is thus tailored to describe the ground state.

A further sophistication consists in allowing the nucleons to couple to angular momenta $J=2,4,6,\ldots$, leading to:
\begin{equation}
\begin{split}
H &= \sum_{j,m>0}\epsilon_{j} n_{jm} -\frac{1}{4} \sum_J\sum_{jj'} G^J_{jj'} \times \\
&\sum_M \sum_{mm'nn'} C^{JM}_{jmjn} C^{JM}_{j'm'j'n'} ~ a^{\dagger}_{jm} a^{\dagger}_{jn}a_{j'n'} a_{j'm'},
\label{eq:HJ}
\end{split}
\end{equation}
now with general Clebsch-Gordan coefficients $C^{JM}_{jmjn}$, allowing for ``rotational bands'' of pair excitations with $J>0$.

By adding more and more terms, we go down a hierarchy of models, eventually reaching the generic case
\begin{equation}\begin{split}
H = \sum_{k}\epsilon_{k} n_{k} + \frac{1}{4} \sum_{klmn} V_{klmn} a^{\dagger}_k a^{\dagger}_l a_n a_m.
\label{eq:Hgeneric}
\end{split}\end{equation}
Finding the ground state of these models in the full Hilbert space~\cite{Liu_etal_2020} soon becomes exponentially difficult and is limited to small systems~\cite{Zelevinsky_Volya_2003,Volya_Brown_Zelevinsky_2001}. We discuss approaches to get past this barrier.

\paragraph{Mean field}
A cheap and popular approach is mean-field BCS (Bardeen-Cooper-Schrieffer) theory~\cite{Belyaev1959,50yearsBCS,Sandulescu_Bertsch_2008,Bertsch_2009}, but has notable shortcomings: Apart from inaccurate ground-state energies, it breaks particle number conservation and predicts a phase transition which is absent in finite-size nuclei~\cite{Zelevinsky_Volya_2003}. Some improvement of the BCS theory is possible~\cite{Bayman_1960,Sambataro_2012}.

\paragraph{QMC} Configuration-space quantum Monte Carlo (QMC) techniques~\cite{Cerf_Martin_1993,Koonin_1997A,Koonin_1997B,Nakada_1997,Pieper_2001,Mukherjee_2011,Alhassid_2008,Alhassid_2012,Lingle_Volya_2015} use stochastic probing of the Hilbert space. The largest hindrance to the application of Monte Carlo is the infamous sign problem, which is less pronounced for nuclei. However, it may still appear for general Hamiltonians and requires workarounds~\cite{Koonin_1997A,Koonin_1997B,Pieper_2001,Alhassid_2012}. Statistical errors may grow unfavorably, in Ref.~\cite{Mukherjee_2011} as $L^3$, reaching the order of $1~\text{MeV}$ even for relatively simple models. We also note that early applications of Monte Carlo dealt with constant pairing couplings $G_{jj'}\equiv G$, a limitation that was addressed only relatively recently~\cite{Lingle_Volya_2015}.

\paragraph{DMRG} The density matrix renormalization group (DMRG)~\cite{White_1992,Schollwoeck_2005,Schollwoeck_2011} approximates the wavefunction by truncating the Hilbert space to a subset of relevant states. This truncation exploits the entanglement properties of system (see Section~\ref{sec:entanglement}) and an essentially numerically exact solution for large systems can often be obtained with low effort. This is known to be the case for basic nuclear toy models~\cite{Dukelsky1999,Dukelsky2001,Dukelsky2002}, as well as for model~\eqref{eq:H_Richardson}, which has been investigated in great detail for superconducting grains~\cite{Dukelsky1999grains,Dukelsky2000grains,vonDelft2001,Gobert2004grainsA,Gobert2004grainsB}.
For the case of generic couplings, such as the \textit{pfg}-shell with up to 12 protons and neutrons~\cite{Papenbrock2005,Pittel2006,Thakur2008,Legeza2015,Tichai2022}, DMRG starts to struggle. A recent algorithmic improvement has been to use an adaptive single-particle basis to reduce entanglement~\cite{Legeza2015,Tichai2022}.

In this paper, we want to study how DMRG performs as the model is tuned to be more and more realistic. This means that instead of jumping to the completely generic case~\eqref{eq:Hgeneric}, we steadily increase the complexity and look at the intermediate cases of~\eqref{eq:H} and~\eqref{eq:HJ}, motivated by its good performance for~\eqref{eq:H_Richardson}. A common playground in this context are the magic-shell nuclei, where one can assume spherical symmetry and neglect the interactions between the two nucleon species, solely focusing on the filling of either the neutron or the proton shell. Since Pb is the largest system among these, it lies at the center of our investigation.

\section{\label{sec:model}The Model}

We investigate the models~\eqref{eq:HJ} and~\eqref{eq:H} (the latter results from~\eqref{eq:HJ} by only keeping the $J=0$ term with $G_{jj'}=G^{J=0}_{jj'}$) which have two inputs: 
(i) The single-particle parameters leading to a particular level structure $\epsilon_j$, for which we use the Seminole parametrization~\cite{Schwierz_etal_2007} of the Woods-Saxon potential;
(ii) the interaction matrix elements $G_{jj'}$ that have to be computed for an effective internucleon interaction. Here, we employ the ``delta-force'' or ``contact interaction'' $V\lr{\vec{r}_1-\vec{r}_2}=V_0~\delta^{\lr{3}}\lr{\vec{r}_1-\vec{r}_2}$, which is a reasonable first guess~\cite{Dobaczewski_1995,Dobaczewski_1996,Burglin_Rowley_1997,Sandulescu_Bertsch_2008,Armstrong_2012,Talmi_1993,Stepanov_2018}:
\begin{equation}\label{eq:contactint}
G^{J}_{jj'} = V_0^J~\matrixel{jj;JM}{\delta^{\lr{3}}\lr{\vec{r}_1-\vec{r}_2}}{j'j';JM}.
\end{equation}
This expression is evaluated using the single-particle solution $\ket{jj;JM} \allowbreak = \allowbreak \frac{1}{\sqrt{2}}\sum_{mm'} \allowbreak C^{JM}_{jmjm'} \allowbreak a^{\dagger}_{jm} a^{\dagger}_{jm'} \ket{0}$, where $\ket{0}$ is the vacuum state (see the Supporting Information for more details). For model~\eqref{eq:H}, $V_0=V_0^{J=0}$ remains as the only free parameter. The procedure to fix the parameters for model~\eqref{eq:HJ} is discussed in Section~\ref{sec:Jcoupling}.

In model~\eqref{eq:H}, there is no term that breaks the pairs, so that the corresponding states can be disregarded, facilitating the computations. We can do this by transitioning to pseudospin operators $L^{\pm}_j$~\cite{Anderson_1958}:
\begin{equation}
\begin{split}
L^{+}_j =  \lr{L^{-}_j}^{\dagger} &= \frac{1}{2}\sum_{m=-j}^j \lr{-1}^{j-m} a^{\dagger}_{jm} a^{\dagger}_{j,-m} \\
&= \sum_{m>0} \lr{-1}^{j-m} a^{\dagger}_{jm} a^{\dagger}_{j,-m}\\
L^z_j &= \frac{1}{2} \sum_{m=-j}^j \lr{a^{\dagger}_{jm}a_{jm}-\frac{1}{2}}\\
      &= \frac{1}{2} \sum_{m>0} \lr{a^{\dagger}_{jm}a_{jm}+a^{\dagger}_{j,-m}a_{j,-m}-1}\\
      &= \frac{1}{2}\lr{n_j-\Omega_j},
\end{split}
\end{equation}
with $\Omega_j=\lr{2j+1}/2$. They fulfill the SU(2) spin algebra relations $\big[L^{+}_j,L^{-}_{j'}\big]=2\delta_{jj'}L^z_j$ and $\big[L^{z}_j,L^{\pm}_{j'}\big]=\pm\delta_{jj'}L^{\pm}_j$.
With these, the Hamiltonian~\eqref{eq:H} can be rewritten as
\begin{equation}
H = \sum_{j}\epsilon_{j} \lr{2L^z_j+\Omega_j} - \sum_{jj'} G_{jj'}L^{+}_j L^{-}_{j'}.
\end{equation}
Furthermore, it will be useful to introduce the total pseudospin operators:
\begin{equation}
L^{\pm,z}_{\text{tot}} = \sum_jL^{\pm,z}_j.
\label{eq:Ltot}
\end{equation}
Particle number conservation translates to the conservation of $L^z_{\text{tot}}$.

\section{\label{sec:dmrg} Applying DMRG to the nuclear pairing Hamiltonian}

\subsection{Matrix-product states; mapping to a chain}

The DMRG algorithm provides a way to compress the many-body wavefunction~\cite{White_1992,Schollwoeck_2011}. By splitting the system into $L$ sites, each with its local basis $\ket{\sigma_{l=1\ldots D}}$ of size $D$, one can transform the basis state coefficients $c_{\sigma_1\sigma_2\ldots\sigma_L}$ into a product of matrices:
\begin{equation}
\begin{split}
\ket{\Psi} &= \sum_{\boldsymbol{\sigma} } c_{\sigma_1\sigma_2\ldots\sigma_L}  \ket{\sigma_1}\ket{\sigma_2}\ldots\ket{\sigma_L} \\
           &= \text{Tr} \sum_{\boldsymbol{\sigma} }\underline{A}^{\sigma_1} \underline{A}^{\sigma_2} \ldots \underline{A}^{\sigma_L} \ket{\sigma_1}\ket{\sigma_2}\ldots\ket{\sigma_L},
\label{eq:MPS}
\end{split}
\end{equation}
where we use open boundary conditions, so that $\underline{A}^{\sigma_1}$ ($\underline{A}^{\sigma_L}$) always has one row (column). The key advantage of such an MPS representation is the notion of {\it locality}: The matrix $\underline{A}^{\sigma_l}$ associated with the site $\ket{\sigma_l}$ depends on $\sigma_l$ only, and this is the very reason why states with certain entanglement properties can be efficiently encoded (see the next subsection). The concrete choice of the `site' $\ket{\sigma_l}$ is in principle arbitrary and determined by practicality (it can be as small as a single orbital).

\begin{figure}
\centering
\includegraphics[width=0.85\columnwidth]{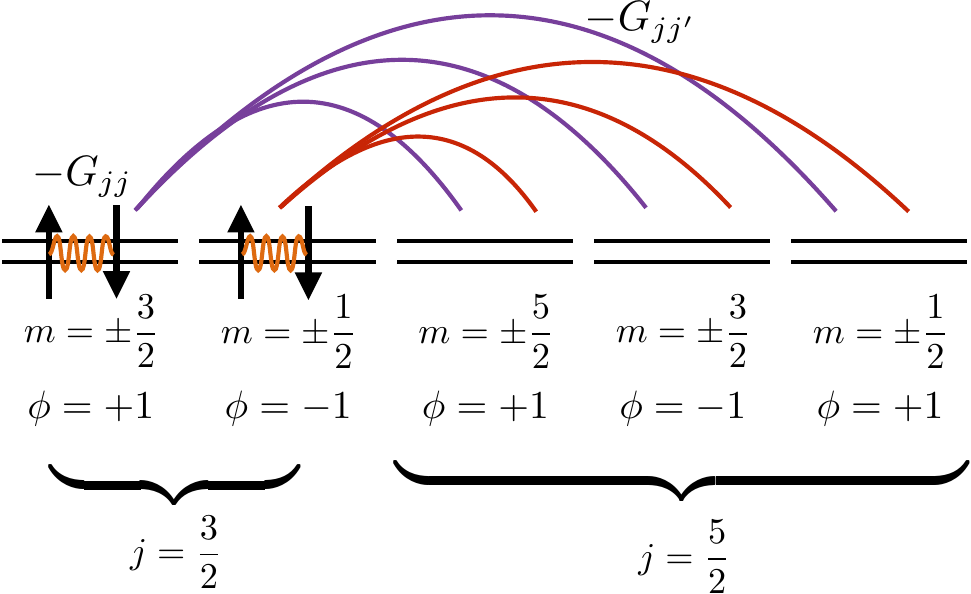}
\caption{
The nuclear pairing Hamiltonian in Equation~(\ref{eq:H}) can be mapped to a 1D chain, which can subsequently be treated using matrix-product states. This is illustrated here for the case of two nuclear levels with $j=\frac{3}{2}$ and $j=\frac{5}{2}$, which are mapped onto a chain of length $L=5$. The arched lines indicate the long-range pair hopping $-G_{jj'}$ between the sites with an additional phase $\phi=\lr{-1}^{j-m}$ ($m>0$) for each site.
}
\label{fig:1_jlevels}
\end{figure}

In our case, it is convenient to group the time-conjugate Kramers pairs with $\pm m$ (where $m>0$) into one site, with the local basis given by the following $D=4$ states: the vacuum (pseudospin-down) ${\ket{\Downarrow}_{jm}}\allowbreak=\allowbreak{\ket{0}}$, the singly occupied states ${\ket{\uparrow}_{jm}}\allowbreak={a^{\dagger}_{jm}}{\ket{0}}$ as well as ${\ket{\downarrow}_{jm}}\allowbreak=\allowbreak{a^{\dagger}_{j,-m}}{\ket{0}}$, and the doubly occupied (pseudospin-up) state ${\ket{\uparrow\downarrow}_{jm}}\allowbreak=\allowbreak{\ket{\Uparrow}_{jm}}\allowbreak=\allowbreak{a^{\dagger}_{jm}}{a^{\dagger}_{j,-m}}{\ket{0}}$. The factor $\lr{-1}^{j-m}$ is absorbed into the Hamiltonian. Thus, a $j$-shell corresponds to $\Omega_j=j+1/2$ sites with energy $\epsilon_j$. All of the sites are sequentially enumerated, so that the whole problem is mapped onto a 1D chain of length $L=\sum_j \Omega_j$. This is illustrated in Figure~\ref{fig:1_jlevels}.

\subsection{\label{sec:entanglement}Entanglement entropy}

Equation~(\ref{eq:MPS}) is exact if the matrices are not truncated; in order to represent an arbitrary state, exponentially large matrices of size up to $D^{L/2}$ are necessary. At the heart of the DMRG algorithm lies the observation that a large class of {\it physical states} can in fact be expressed by matrices of a much smaller size $\chi \ll D^{L/2}$ (called the `bond dimension') without introducing much error. The value $\chi$ is a measure of the number of variational parameters that are used to represent the wavefunction. It is also a measure of the numerical effort, since the largest matrix that must be handled is of size $\chi\times\chi$.

The reason why many physical states can be efficiently represented by an MPS is related to the notion of entanglement. To quantify this, one divides the chain into two parts A and B and integrates out part B, resulting in a mixed state:
\begin{equation}
\rho_A = \text{Tr}_B \ket{\Psi}\bra{\Psi}.
\end{equation}
The quantity that characterizes the entanglement of this bipartition is the entanglement entropy
\begin{equation}
S_{A|B} = -\text{Tr}_A\rho_A \ln \rho_A,
\label{eq:S}
\end{equation}
which is granted as a byproduct in the DMRG algorithm. The crucial question is thus how $S_{A|B} $ scales with the number of particles. For gapped 1D models with short-range interactions, the so-called area law guarantees that $S_{A|B}$ stays constant~\cite{Eisert_2010}, so that the numerical effort grows only linearly.

The nuclear pairing problem features long-range interactions, and the existence of an area law is not guaranteed. Even though the system is finite, the amount of entanglement (and thus $\chi$) might still be prohibitively large. However, we find that the ground state has very low entanglement and can be represented by a rather small $\chi \sim 10^2$ to nearly machine precision. The reason is that we are in the weak-coupling regime and the deep $j$-shells in the left part of the chain are almost completely filled, forming a weakly entangled near-product state, while the right part of the chain is mostly empty, with only few excitations across a softened Fermi edge that contribute to the entanglement. This will be demonstrated explicitly in Section~\ref{sec:Pb}.

Finally, we emphasize that the DMRG approach does not neglect any diagrams and does not rely on particle-hole excitation cutoffs as, e.g., in the configuration interaction approach~\cite{ComputationalNuclear2017}. If a good MPS representation of the ground state can be found, the result becomes numerically exact.

\subsection{Matrix-product operators}

In order to implement the DMRG algorithm, we also need to express the Hamiltonian in a form analogous to the one of the wavefunction in Equation~(\ref{eq:MPS}). This is achieved by a matrix-product operator (MPO) representation:
\begin{equation}
H = \sum_{\boldsymbol{\sigma}\boldsymbol{\sigma}'} \text{Tr}~\underline{W}^{\sigma_1\sigma_1'} \ldots \underline{W}^{\sigma_L\sigma_L'}
\ket{\sigma_1}\ldots\ket{\sigma_L} \bra{\sigma_1'}\ldots\bra{\sigma_L'}
\label{eq:MPO}
\end{equation}
which has some MPO bond dimension $\chi_{\text{MPO}}$.
Having chosen the sites (see Figure~\ref{fig:1_jlevels}), any local operator is exactly encodable with $\chi_{\text{MPO}}=1$, while a sum of local terms such as Equation~\eqref{eq:Ltot} is encodable with $\chi_{\text{MPO}}=2$~\cite{Schollwoeck_2011}. There are simple formulas for adding (multiplying) terms on the MPO level~\cite{Schollwoeck_2011}, whereby the bond dimensions add (multiply). The resulting $\chi_{\text{MPO}}$ of this naive approach can in general be prohibitively large for many long-range terms. Fortunately, this representation is not unique and it is possible to apply a {\it lossless} MPO compression algorithm~\cite{Hubig_2017} that exploits linear dependencies in the matrices to reduce $\chi_{\text{MPO}}$. For the given problems, we find a reduction by as much as $99\%$ from the naive construction; the runtime of this algorithm is a few seconds. No approximation is introduced by this compression procedure.

The massive reduction of $\chi_{\text{MPO}}$ is probably a specific feature of the nuclear pairing Hamiltonian and likely related to the fact that the interaction terms are the same for all sites between fixed $j$ and $j'$. We suspect that the efficient representation can also be found analytically in the case of singlet pairing. Note that in the Richardson model~\eqref{eq:H_Richardson}, where $G_{jj'}\equiv G$, only the total pseudospinflip operators in Equation~(\ref{eq:Ltot}) couple to each other in the form $-GL^{+}_{\text{tot}}L^{-}_{\text{tot}}$ and are at most described by $\chi_{\text{MPO}}=4$~\cite{Pirvu_2010}. However, if more complicated terms are added to the Hamiltonian (like the pairing to $J>0$ considered in Section~\ref{sec:Jcoupling}), analytical approaches become too cumbersome and the numerical compression that we use becomes indispensable.

\subsection{Implementation}

The core idea of a ground-state DMRG algorithm is to locally optimize the tensors $\underline{A}^{\sigma_i}$ from Equation~(\ref{eq:MPS}) using the Ritz variational principle \cite{White_1992,Schollwoeck_2005,Schollwoeck_2011}. In the presence of symmetries the tensors acquire a block structure and the algorithm has to distribute the available bond dimension among these blocks. For this, we use the subspace expansion method of the 1-site algorithm described in Ref.~\cite{Hubig_2015}.

As a measure of convergence we use the energy variance per particle
\begin{equation}
\Delta E^2/N=\lr{\avg{H^2}-E^2}/N,
\label{eq:var}
\end{equation}
which indicates how close we are to an eigenstate.

In condensed-matter language, the Hamiltonian in Equation~(\ref{eq:H}) is mapped to a chain of fermionic orbital pairs with on-site energies $\epsilon_j$, an on-site attractive Hubbard term $-G_{jj}n_{jm}n_{j,-m}$, and attractive pair hoppings between all sites (see Figure~\ref{fig:1_jlevels}). Since the single-band Hubbard model is a standard application for the DMRG method, we point out that the code can be implemented without any significant modification of standard algorithms, with the exception of the MPO representation of the Hamiltonian (see above). Another obstacle is the impeded convergence due to the slightly pathological nature of the Hamiltonian~\eqref{eq:H} (but not of~\eqref{eq:HJ}). We discuss how it can be resolved in Appendix~\ref{app:stuck}.

\section{\label{sec:benchmark} Benchmark: tin isotopes}

\begin{figure}
\centering
\includegraphics[width=\columnwidth]{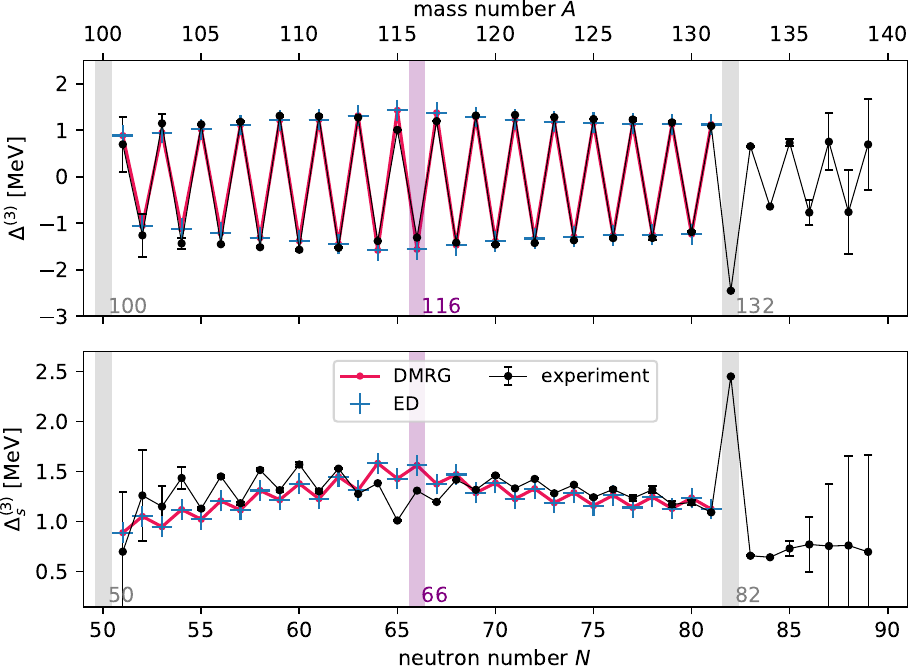}
\caption{
The Sn isotopes serve as a benchmark case for the DMRG algorithm. The even-odd mass differences~(\ref{eq:Delta3}) as well as the staggered variant (\ref{eq:Delta3s}) are compared to experimental data~\cite{AME2020} and exact diagonalization (ED) results; the latter are reproduced by the DMRG to 10 relevant digits. The single-particle levels $\epsilon_j$ as the interaction matrix elements $G_{jj'}$ are taken from Ref.~\cite{Zelevinsky_Volya_2003}. The grey and purple shaded areas indicate closed shells and half filling, respectively.
}
\label{fig:gapsSn}
\end{figure}

It is instructive to study a drosophila scenario to benchmark the DMRG algorithm. To this end, we consider pairing in the Sn isotopes, which can still be treated using exact diagonalization \cite{Holt_1998} and which frequently serve as a testing ground for approximate techniques~\cite{Zelevinsky_Volya_2003,Sambataro_2012,Claeys_2018}. 
There is a given set of matrix elements for $\epsilon_j$ and $G_{jj'}$~\cite{Zelevinsky_Volya_2003}, which we use in this case (in all other cases we compute them as described in Section~\ref{sec:model}).
The system is assumed to have an inert core with $Z=50$ protons and $N=50$ neutrons. The valence neutrons fill the shells $1g_{7/2}$, $2d_{5/2}$, $2d_{3/2}$, $3s_{1/2}$, and $1h_{11/2}$ from $N=50$ to $N=82$. In our nomenclature, this corresponds to a chain of $L=16$ sites. To fix the filling, we only exploit the conservation of the particle number (U(1) charge symmetry).

The MPO associated with the Hamiltonian of the Sn isotopes can be compressed exactly in under 0.1s runtime from the maximal bond dimension of $\chi_{\text{MPO}}=130$ down to $\chi_{\text{MPO}}=8$. The ground state search is most difficult around half filling ($N=66$ or 16 fermions in the chain), but the runtime is still less than 1s on a regular desktop computer. The MPS bond dimension is at most $\chi\sim 50$. The energy variance per particle (Equation~(\ref{eq:var})) is of the order of $10^{-9}~\text{MeV}^2$ or smaller, so that we can be confident to have well-converged results.

A quantity that demonstrates the even-odd effect due to pairing is the discrete derivative of the ground-state energy with respect to the particle number, which we compute via the 3-point rule~\cite{Bender_2000}
\begin{equation}
\Delta^{\lr{3}}\lr{N} = E_0\lr{N}-\left[E_0\lr{N-1}+E_0\lr{N+1}\right]/2.
\label{eq:Delta3}
\end{equation}
This is known as the `even-odd mass difference'. One can also add a staggered factor that makes all values positive:
\begin{equation}
\Delta_s^{\lr{3}}\lr{N} = \lr{-1}^{N+1} \Delta^{\lr{3}}\lr{N}.
\label{eq:Delta3s}
\end{equation}
The result is shown in Figure~\ref{fig:gapsSn}; exact diagonalization data are reproduced to at least 10 digits. We also show experimental results ~\cite{AME2020}. The agreement between theory and experiment is noticeably worse around half filling, where there is a gap between $2d_{5/2}$ and $3s_{1/2}$ in the single-particle energies.

\begin{figure}
\centering
\includegraphics[width=\columnwidth]{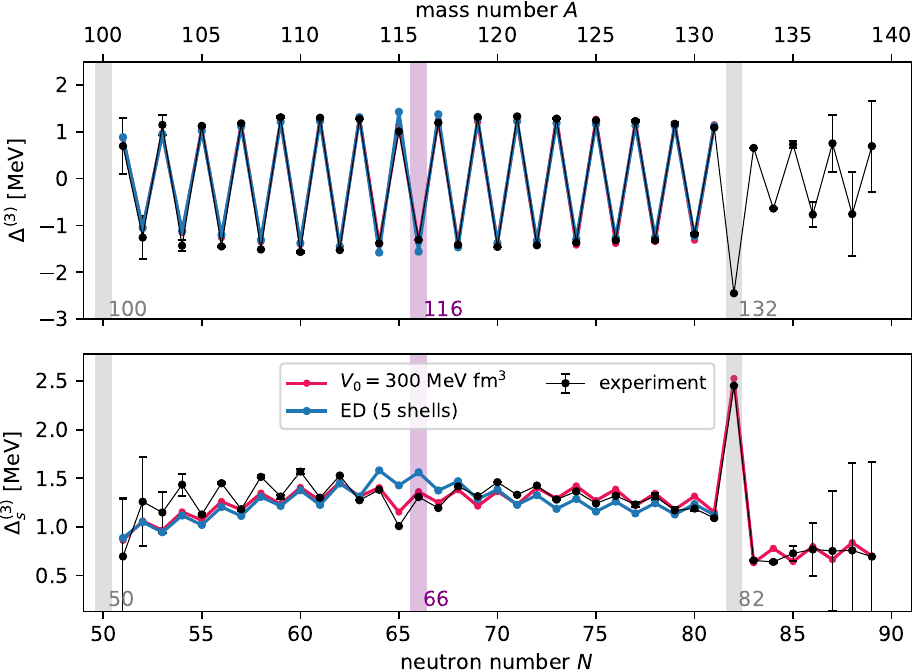}
\caption{
Even-odd mass differences of Sn calculated with the DMRG for a large configuration space that includes 10 $j$-shells between the magic numbers $N=50$ and $N=126$ (all bound eigenstates). The matrix elements $\epsilon_j$ and $G_{jj'}$ are determined from the single-particle solution (see Section~\ref{sec:model}) with $V_0$ being the only free parameter. The agreement between theory and experiment is significantly better than the typically used benchmark dataset in Figure~\ref{fig:gapsSn}, where only five shells were included (`ED' shows the same dataset as in Figure~\ref{fig:gapsSn}).
}
\label{fig:gapsSnDMRG}
\end{figure}

We try to improve the agreement with the experimental data by extending the parameter space from five shells to all bound eigenstates above $N=50$. We add $2f_{7/2}$, $3p_{3/2}$, $3p_{1/2}$, $1h_{9/2}$, and $2f_{5/2}$ ($1i_{13/2}$ turns out to be unbound) to the configuration space, resulting in a chain of $L=31$ sites; and use the procedure described in Section~\ref{sec:model} to compute $\epsilon_j$ and $G_{jj'}$. The result is shown in Figure~\ref{fig:gapsSnDMRG} for the optimal $V_0=300~\MeVfm$. The average deviation from the experimental values has now decreased to $\sigma\approx 0.10~\text{MeV}$. In particular, the dip around half filling is captured more accurately. Moreover, we can compare with the neutron-rich isotopes up to $^{140}$Sn that were not considered previously using ED, and still find excellent agreement with the experiment~\cite{AME2020}.

\section{\label{sec:Pb} The lead isotopes}

\begin{figure*}
\centering
\includegraphics[width=0.7\textwidth]{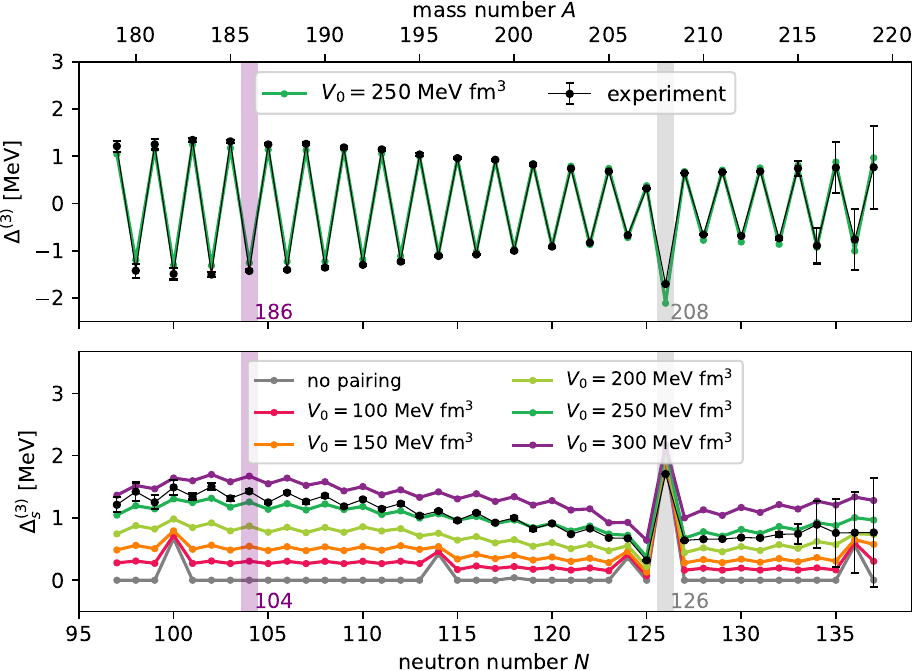}
\caption{
The even-odd mass differences [see Eqs.~(\ref{eq:Delta3}) and (\ref{eq:Delta3s})] are calculated using the DMRG for the lead isotopes and compared to the experimental data (black). We use a large configuration space that includes all shells between $N=82$ and $N=184$. The matrix elements $\epsilon_j$ and $G_{jj'}$ are determined from the single-particle solution (see Section~\ref{sec:model}) with $V_0$ being the only free parameter. For $V_0=250~\text{MeV}$, we find excellent agreement with the experimental data~\cite{AME2020}.
}
\label{fig:gapsPb}
\end{figure*}

We now turn to the more challenging Pb isotopes, which have a very large Hilbert space around half filling. We assume an inert core of $Z=82$ protons and $N=82$ neutrons and include all the shells between $N=82$ and $N=126$ into our parameter space ($2f_{7/2}$, $1h_{9/2}$, $1i_{13/2}$, $3p_{3/2}$, $2f_{5/2}$, $3p_{1/2}$), as well as all the shells up to the next magic number $N=184$ ($2g_{9/2}$, $1i_{11/2}$, $1j_{15/2}$, $3d_{5/2}$, $4s_{1/2}$, $2g_{7/2}$, $2d_{3/2}$). This is a very large configuration space that corresponds to a chain length of $L=51$. The half filling point of the $N=82-126$ major shell is given by $N=104$ ($^{186}$Pb), or by 22 fermions in the chain. The smallest and largest known isotopes are $^{178}$Pb and $^{220}$Pb~\cite{AME2020}, respectively (corresponding to 14 and 56 fermions in the chain). The matrix elements $\epsilon_j$ and $G_{jj'}$ are determined from the single-particle solution (see Section~\ref{sec:model}). Note that large-scale ED shell model calculations have been performed for this system for doping of up to 14 neutrons away from the magic $N=126$ shell, i.e. $\Delta^{\lr{3}}$ has been computed in a range of $N=114-126$~\cite{Qi_Jia_Fu_2016}.

The MPO representation of the Hamiltonian for this much larger system can be again compressed from the naive setup that results in $\chi_{\text{MPO}}=1302$ down to $\chi_{\text{MPO}}=16$. An MPS bond dimension of around $\chi\sim 250$ is sufficient to faithfully represent the ground state at intermediate fillings, and the variance per particle~(\ref{eq:var}) is again at worst of the order of $10^{-9}~\text{MeV}^2$. The calculation takes about 1-2 minutes on a desktop computer.

The results for the even-odd mass difference $\Delta^{\lr{3}}$ are displayed in Figure~\ref{fig:gapsPb}. We find that a pairing strength of $V_0=250~\MeVfm$ yields excellent agreement with the experimental data~\cite{AME2020} despite the simplicity of the delta-force interaction.
This can be quantified by introducing
\begin{equation}
\sigma =  \sqrt{ \frac{\sum_{N=N_{\text{min}}}^{N_{\text{max}}} \lr{\Delta^{\lr{3}}_{\text{theory}}\lr{N}-\Delta^{\lr{3}}_{\text{exper.}}\lr{N}}^2} {N_{\text{max}}-N_{\text{min}}}},
\label{eq:err}
\end{equation}
and we find $\sigma\approx 0.13~\text{MeV}$ (for the Sn benchmark, the corresponding value is $\sigma\approx 0.16~\text{MeV}$). This should be compared to the scale of $\Delta^{\lr{3}}$, which is of the order of $1~\text{MeV}$. In Appendix~\ref{sec:BCS}, we also assess the accuracy of BCS in computing $\Delta^{\lr{3}}$ for Sn and Pb.

Our approach allows us to look at fillings in higher shells that are normally left out of the active space. This is discussed in the Supporting Information.

\section{\label{sec:S} Entanglement entropy}

\begin{figure}
\centering
\includegraphics[width=\columnwidth]{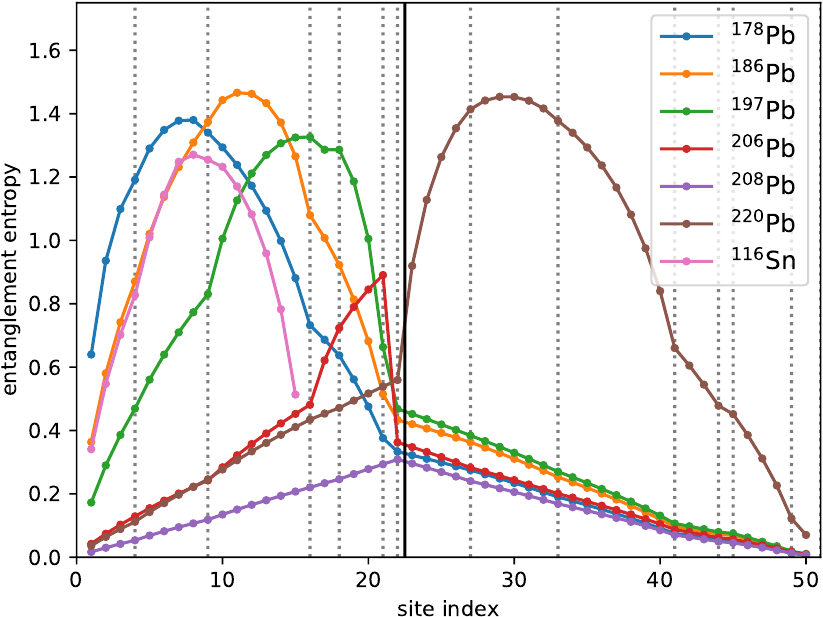}
\caption{
Entanglement entropy (\ref{eq:S}) for different bipartitions on the chain geometry (the lower single-particle orbitals are to the left) for various Pb isotopes as well as for the half-filled $^{116}$Sn. Closures of the various $j$-shells of Pb in the single-particle picture are indicated by the dotted vertical lines. The Pb magic gap is indicated by the solid vertical line.
}
\label{fig:S}
\end{figure}

For both Sn and Pb, we compute the entanglement entropy given by Equation~(\ref{eq:S}) between the lower and higher single-particle orbitals, or equivalently, between the left and right part of the chain. The result is shown in Figure~\ref{fig:S}. Not surprisingly, the largest values of $S_{A|B}\approx1.4$ are found for open-shell isotopes $^{186}$Pb and $^{220}$Pb, while the closed-shell $^{208}$Pb only has maximal value of around $S_{A|B}\approx0.3$. Most importantly, we see that the nuclear shell structure acts as a natural `entanglement barrier': Beyond half filling, the shell starts to close again and the entanglement decreases as the wavefunction gets closer to a product state. Since entanglement depends on the choice of basis, we conclude that the Woods-Saxon basis of the effective shell models is already a very good choice to minimize it. Coupled with the possibility to represent Hamiltonians \eqref{eq:H_Richardson}, \eqref{eq:H}, and \eqref{eq:HJ} by only moderately-sized MPOs, this explains the very good performance of DMRG for these systems. Note that the decrease of $S_{A|B}$ has been proposed as a measure to detect emergent shell closures in the generic case~\cite{Tichai2022}.

In Appendix~\ref{sec:spec}, we additionally test the capabilities of the DMRG method to access many excited states and spectral functions.

\section{\label{sec:Jcoupling} Coupling to finite angular momentum}

As a final point, we demonstrate that the DMRG is also capable of effectively treating Hamiltonians that are more general than the pure singlet pairing and investigate model~\eqref{eq:HJ} with contact interaction~\eqref{eq:contactint}.

For singlet pairing, it was sufficient to exploit particle number conservation only. At this point, we exploit the conservation of the angular momentum projection $J^z$ as well. For every multiplet with integer $J=0,2,4,6,8$ (each containing $2J+1$ states), there is now exactly one state that lies in the sector with $\avg{J^z}=0$. We can determine $J$ by directly computing $\avg{\vec{J}^2}=J\lr{J+1}$.

The MPO bond dimension is now significantly larger than before due to the presence of many long-range terms. Restricting ourselves to the $N=82-126$ ($N=126-184$) major shell, the MPO can be compressed down to $\chi_{\text{MPO}}=166$ ($\chi_{\text{MPO}}=234$). 

A textbook case are the proton excitations of $^{210}$Po, which form a sequence with total angular momenta of $J=2,4,6,8$. We extend this analysis to higher isotones by continuing to fill the proton levels up to $^{216}$Th. In contrast to the literature, we do not only fill the lowest $1h_{9/2}$ level, but take into account the full space made up of $1h_{9/2}$, $2f_{7/2}$, $1i_{13/2}$, $2f_{5/2}$, and $3p_{3/2}$ ($L=21$). For these moderate proton fillings, the variance per particle is around $10^{-8} \text{MeV}^2$ or less, and the runtime is of the order of tens of minutes.

We determine the parameters $V_0^{J=0}$ by fitting the even-odd mass differences of the pure singlet pairing Hamiltonian (\ref{eq:H}) to the experimental data, while the remaining parameters are fitted to the measured splitting of the four lowest eigenenergies of $^{210}$Po. We obtain, in units of $\MeVfm$, $V_0^{J=0}=320$, $V_0^{J=2}=360$, $V_0^{J=4}=230$, $V_0^{J=6}=220$, and $V_0^{J\geq8}=-210$.
 
\begin{figure}
\centering
\includegraphics[width=\columnwidth]{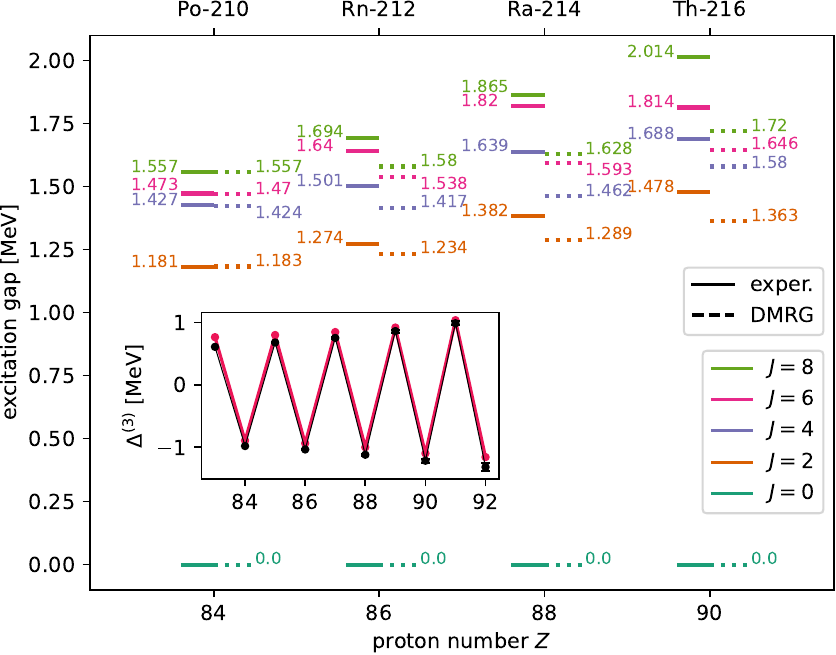}
\caption{
Excitation energies of the $N=126$ isotones above $Z=82$ calculated with the Hamiltonian (\ref{eq:HJ}) that goes beyond singlet pairing. The solid and dotted lines are experimental values and DMRG results, respectively. The inset shows the even-odd mass differences up to $^{218}$U for $V_0^{J=0}=320~\MeVfm$ (red) in comparison with the experimental values (black)~\cite{AME2020}.
}
\label{fig:levelsNeu126}
\end{figure}

The spectra of the higher isotones up to $^{216}$Th are shown in Figure~\ref{fig:levelsNeu126} (the inset displays the even-odd mass differences used to determine $V_0^{J=0}$). While the increase of the energies relative to the ground state is captured qualitatively, the splitting is underestimated quantitatively and clearly requires using a better effective interaction~\cite{Stepanov_2018,Gottardo_2012}.

\section{\label{sec:conclusion} Conclusion and outlook}

The density-matrix renormalization group is very suitable to study simplified nuclear shell models. We argue that the DMRG method can supersede the mean-field BCS approach for a broad range of problems, bringing advantages across the board. We have shown that it is effective in dealing with generalized models that have level-dependent matrix elements and additional coupling terms. One can obtain very accurate results --- basically up to machine precision --- with modest numerical effort even for a very large number of shells. The reason for the efficiency is the naturally low entanglement in the Woods-Saxon basis.

We have demonstrated the capabilities of the method by calculating the even-odd mass differences for all known isotopes of Pb. We found that experimental results can be reproduced quantitatively by choosing a universal value of $V_0=250~\MeVfm$ across all neutron fillings.

The method does not yet run into significant problems when computing excited states for  pairing with $J>0$, so more terms can still be added to the Hamiltonian. If one goes away from the magic shells, deformation starts to play a role~\cite{Belyaev1959}, but it only enters on the single-particle level (e.g., as a deformed Woods-Saxon potential) and can thus be accounted for easily. Studying finite temperatures, which relates to nuclear thermometry~\cite{Kelic_2006,Sumaryada_Volya_2007}, would be possible within the grand-canonical ensemble by doubling the system~\cite{Feiguin_White_2005,Schollwoeck_2011,karrasch12,karrasch16}. Another interesting application could be the alpha clustering problem~\cite{Norrby2008,Bai2018} as an extension of pairing models.

\medskip
\textbf{Acknowledgements} \par 
We would like to thank Rok Žitko for helpful discussions.\\
C.K. and R.R. acknowledge support by the Deutsche Forschungsgemeinschaft through the Emmy Noether program (Grant No. KA 3360/2-1) as well as from ``Nieders\"achsisches Vorab'' through the “Quantum- and Nano-Metrology (QUANOMET)” initiative within Project No. P-1.
C.P. is supported by the Deutsche Forschungsgemeinschaft (DFG) through the Cluster of Excellence Advanced Imaging of Matter -- EXC 2056 -- project ID 390715994.
M.P. received funding from the European Research Council (ERC) under the European Union’s Horizon 2020 research and innovation programme (Grant Agreement No. 677061).
\medskip


\appendix

\section{\label{app:stuck}Dealing with convergence problems}

The pairing Hamiltonian~\eqref{eq:H} has the peculiarity that for odd particle numbers, an unpaired one remains ``without a partner'' and is unaffected by the pair hopping, thus blocking a particular, \textit{a priori} unknown orbital. Within exact diagonalization, this can be exploited in the `seniority scheme' to block-diagonalize the Hamiltonian for all possible values of the seniority $s_j$, which is essentially the number of unpaired particles in the $j$-shell~\cite{Liu_etal_2020}.

Like any numerical variational approach, the DMRG algorithm is prone to getting stuck in states which are not eigenstates (signalled by a large variance Equation~\eqref{eq:var}) or not the true ground state. In our case, a single particle that gets stuck in an orbital causes the DMRG algorithm to converge to a pair-broken excited eigenstate. This problem can be solved for even fillings where we can safely remove the singly occupied states from the local basis and work only with $\ket{\Downarrow}_{jm}$ and $\ket{\Uparrow}_{jm}$ at each site (which is equivalent to a pseudospin representation).

In order to resolve this issue for odd fillings, we take the full local basis in a given $j$-shell and the restricted basis $\ket{\Downarrow}_{jm},\ket{\Uparrow}_{jm}$ everywhere else; and compute the lowest state for this configuration. The true ground state is given by the one with the lowest energy among all selections of $j$. Still, we find that this procedure tends to get stuck with a large variance (Equation~\eqref{eq:var}). A simple way to help the algorithm is to first determine the ground state $\ket{E_{0}(N)}$ for even fillings $N$ and to subsequently use $\sum_m \allowbreak a^{\dagger}_{jm} \allowbreak \ket{E_{0}(N, \text{even})}$ as initial guesses for odd fillings. These states are already very close to eigenstates, and the remaining convergence is very quick.

An alternative approach to prevent the variational algorithm from getting stuck is to perturb the Hamiltonian, $H\to H+\alpha H_p$, where the perturbation $H_p$ should contain hopping terms of the type $a^{\dagger}_{jm}a_{j'm'}$ that make single particles mobile. By slowly letting $\alpha$ go to zero every few iterations, good convergence can be achieved for even fillings. The calculation for odd filling is then performed analogously by using $\sum_m a^{\dagger}_{jm}\ket{E_{0}(N, \text{even})}$ as an initial guess. This alternative method does not rely on restricting the local basis, which is beneficial if the full wavefunction is needed in subsequent steps. We have tested that both of our approaches yield the same ground state.

We note that the issue does not appear when dealing with the Hamiltonian~\eqref{eq:HJ}, where pairs are also formed across the sites.

\section{\label{sec:spec} Excited states, Two-neutron removal spectral function}

The Hamiltonian (\ref{eq:H}) is tailored to describe the ground state: It only pulls the many-body eigenstates with $J=0$ down in energy but leaves the other eigenstates highly degenerate. However, it is reasonable to investigate to which degree it also captures pair excitations with $J=0$. A particularly interesting system is $^{208}$Pb where such `pair vibrations' arise across the magic gap and can be probed by two-neutron pickup and stripping experiments: A proton that passes by a nucleus can pick up two neutrons and end up as a triton~\cite{Igo_Barnes_Flynn_1970}. Vice versa, two neutrons may be stripped and added to the nucleus~\cite{Alford_1983}. (Note that the condensed-matter electronic equivalents are the \textit{Auger Electron Spectroscopy} and the \textit{Appearance Potential Spectroscopy}~\cite{Weissmann_Mueller_1981,Potthoff_1993,Fukuda_2010,Rausch_Potthoff_2016}.) By choosing $^{210}$Pb and $^{206}$Pb as targets, one can thus specifically study the excited neutron states of $^{208}$Pb. If a pair is taken out above (below) the magic gap of $^{210}$Pb, the nucleus should essentially end up in the ground state (an excited state) of $^{208}$Pb.

Within DMRG, the lowest eigenstates can be constructed by first finding the ground state $\ket{E_0}$ and then altering the Hamiltonian to $\tilde{H} = H+E_p\ket{E_0}\bra{E_0}$ with a suitably high energy penalty $E_p$. This lifts the ground state in energy by $E_p$, and the ground state of $\tilde{H}$ will be the first excited eigenstate $\ket{E_1}$. The procedure can then be iterated.

\begin{figure}
\centering
\includegraphics[width=\columnwidth]{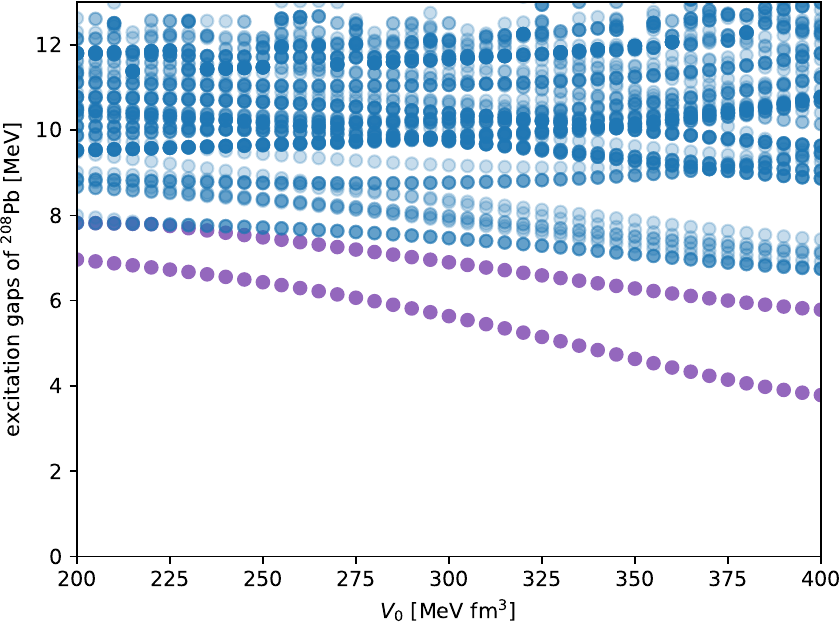}
\caption{
DMRG data for the energies of the first 100 excited states of the pairing Hamiltonian~(\ref{eq:H}) relative to the ground state as a function of the coupling strength $V_0$ for $^{208}$Pb. The lowest two singlet states are marked in purple. In all other cases, a stronger color saturation indicates larger degeneracy.}
\label{fig:Pb208exc}
\end{figure}

Figure~\ref{fig:Pb208exc} shows the first 100 eigenenergies relative to the ground state of the pairing Hamiltonian for $^{208}$Pb as a function of $V_0$ within our parameter space of 13 $j$-shells. As the pairing strength $V_0$ is increased, two non-degenerate singlet states split off from a quasi-continuum of other excitations. This is qualitatively compatible with the experiment where one finds two singlet excitations with $E-E_0=4.868~\text{MeV}$ and $E-E_0=5.241~\text{MeV}$~\cite{NuDat} that lie below a dense sequence of singlets around $E-E_0\approx 8~\text{MeV}$.

We also consider the two-neutron removal spectral function~\cite{Muether_Polls_Dickhoff_1995,Geurts_1996,Amir-Azimi-Nili_1997}:
\begin{equation}
\begin{split}
S^-\lr{E} = \sum_n \big|\matrixel{E_n\lr{N-2}}{L^-_{\text{tot}}}{E_0\lr{N}}\big|^2 \times\\
\delta\lr{E-E_0\lr{N-2}+E_n\lr{N-2}}.
\label{eq:Smtot}
\end{split}
\end{equation}
The shift in the $\delta$-function is such that the ground state $n=0$ contributes at $E=0$.
Within DMRG, Equation~\eqref{eq:Smtot} can be evaluated using a Chebyshev polynomial technique~\cite{KPM_1996,KPM_2006}. The spectral function is decomposed as a sum over Chebyshev polynomials whose coefficients are computed via the Chebyshev recurrence relation $\ket{\Psi_{n+1}}=2H\ket{\Psi_n}-\ket{\Psi_{n-1}}$ starting from the initial state $\ket{\Psi_{0}}=L^-_{\text{tot}} \ket{E_0^{N}}$. Each state $\ket{\Psi_{n}}$ is expressed as an MPS, and the bond dimension is adjusted such that $\lVert \allowbreak\ket{\Psi_{n+1}}\allowbreak-2H\ket{\Psi_n}\allowbreak+\ket{\Psi_{n-1}}\allowbreak\rVert^2 \allowbreak< 10^{-5}$. The resulting energy resolution is approximately given by the many-body bandwidth divided by the number of the Chebyshev moments, $\delta E = \lr{E_{\text{max}}^{N-2}-E_0^{N-2}}/N_{\text{mom}}$. We set $\delta E=0.1~\text{MeV}$, which translates to $N_{\text{mom}} \gtrsim 4000$.

\begin{figure}[t]
\centering
\includegraphics[width=\columnwidth]{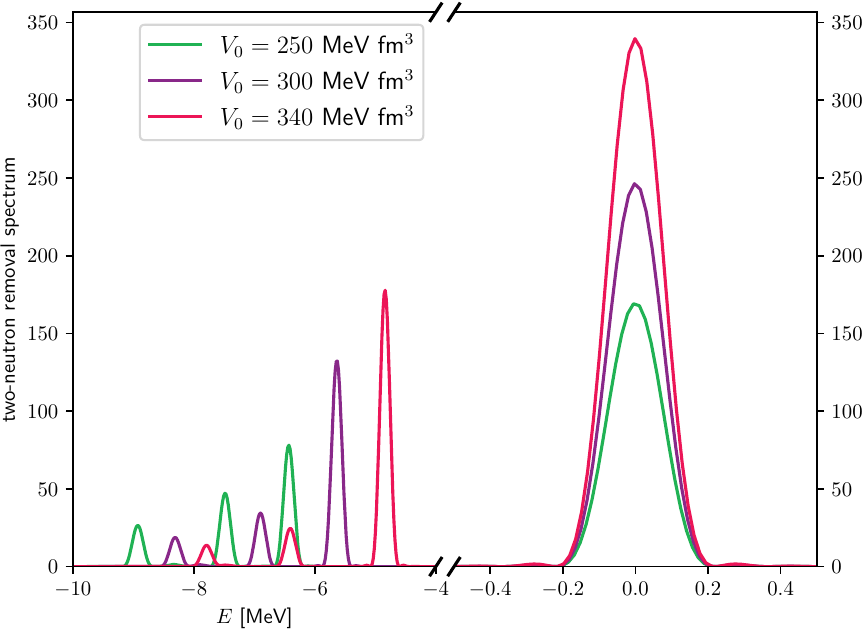}
\caption{
Two-neutron removal spectral function~(\ref{eq:Smtot}), calculated using DMRG for $^{210}$Pb for various pairing strengths at a resolution of $\delta E\approx 0.1~\text{MeV}$.
}
\label{fig:spec}
\end{figure}

The result for $^{210}$Pb is displayed in Figure~\ref{fig:spec}. The peaks are associated with the ground state ($E=0$) as well as with the different excited states. The two peaks closest to $E=0$ correspond to the split-off singlet excitations from Figure~\ref{fig:Pb208exc}. We observe that an increase of the pairing strength results in an increase of spectral weight for the lower singlet and the ground state, but in a reduction for all higher eigenstates.

\section{\label{sec:BCS} Comparison with mean field}

In this section we address the deviations of our approach from mean field theory (BCS), where the pairing terms are replaced by averages and a quadratic Hamiltonian remains. It is known that for the Richardson model, mean field becomes a good approximation for large $L$ (at half filling) or large $G$~\cite{Dukelsky1999grains,vonDelft2001}, as both limits lead to the bulk superconducting gap $\Delta_{\text{BCS}}$ to dominate over the finite-size spacing $d$, i.e. $\Delta_{\text{BCS}} \gg d$.
However, for the model where $G_{jj'}$ are computed from the single-particle solution, it is \textit{a priori} not clear what happens, as a larger nucleus also leads to a weaker coupling because of a smaller radial overlap~\cite{Brink_Broglia_2005}, so that the two effects are to some degree compensatory. Note that the previously obtained optimal coupling value for Pb $V_0=250~\MeVfm$ has be be multiplied with the interaction matrix elements, so that we find that the actual pair hopping strength $G_{jj'}$ is in the range of $0.03-0.29~\text{MeV}$, compared to $0.15-0.72~\text{MeV}$ for Sn.

\begin{figure}
\centering
\includegraphics[width=\columnwidth]{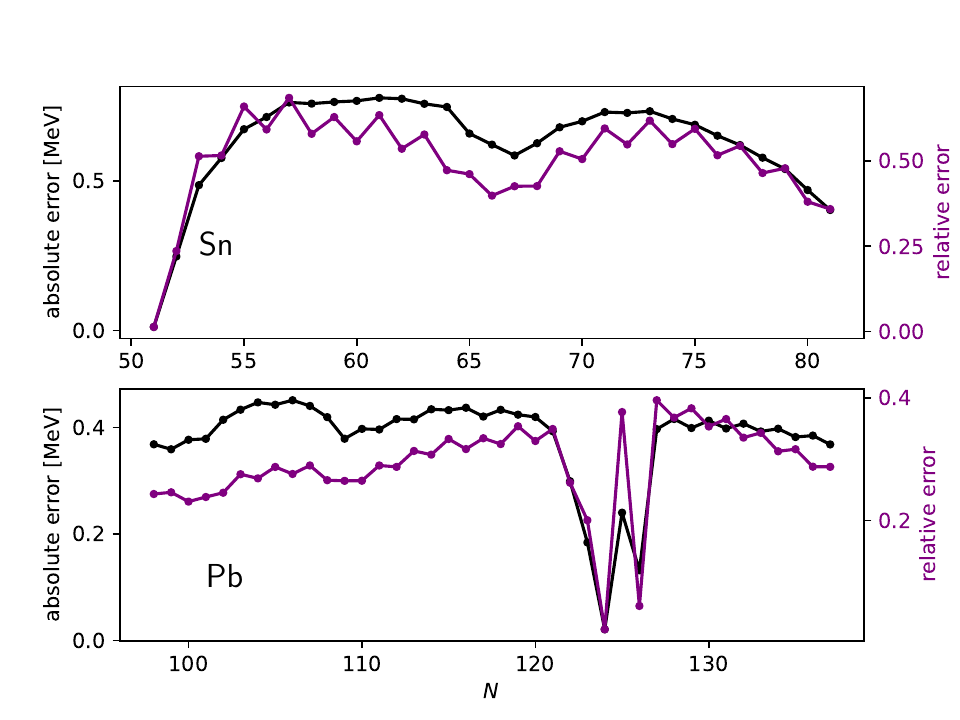}
\caption{
Absolute error $\big|\Delta^{\lr{3}}_{\text{BCS}}-\Delta^{\lr{3}}_{\text{exact}}\big|$ (left scale) and relative error (right scale) $\big|\Delta^{\lr{3}}_{\text{BCS}}-\Delta^{\lr{3}}_{\text{exact}}\big|/\big|\Delta^{\lr{3}}_{\text{exact}}\big|$ of the BCS computation for $V_0=250~\MeVfm$. We take the DMRG result as the exact reference value.
}
\label{fig:BCS}
\end{figure}

Figure~\ref{fig:BCS} shows the deviation of $\Delta^{(3)}$ between the BCS result and the DMRG result (which we take as the exact value). We see that it becomes somewhat smaller for Pb, but still reaches a value of $0.4~\text{MeV}$ in absolute terms, or up to 40\% in relative terms. Thus, BCS does not tie with DMRG in terms of accuracy even for large nuclei.

\clearpage

\setcounter{figure}{0} 
\renewcommand{\thefigure}{S\arabic{figure}} 

\section*{Supporting Information}
\subsection{\label{sec:Pb_n} Occupation numbers for Pb}

\begin{figure}
\centering
\includegraphics[width=\columnwidth]{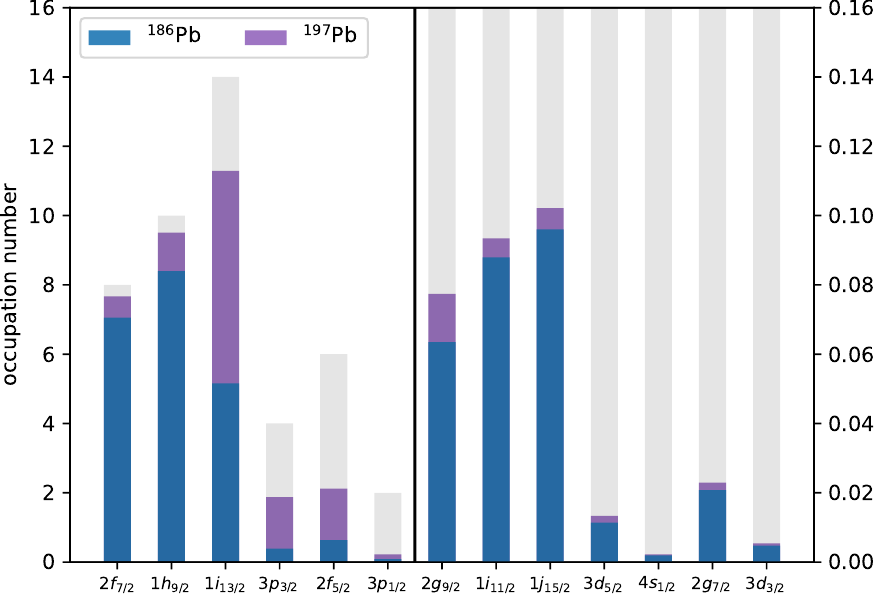}
\caption{
Occupation numbers $\avg{n_j}$ for $^{186}$Pb and $^{197}$Pb at $V_0=250~\MeVfm$ as computed with DMRG. The scale of the y-axis for the levels below (above) the magic number $N=126$ is shown on the left (right) side. The location of the gap is marked by the vertical line and grey shading indicates the maximum shell capacity.
}
\label{fig:n}
\end{figure}

\begin{figure}
\centering
\includegraphics[width=\columnwidth]{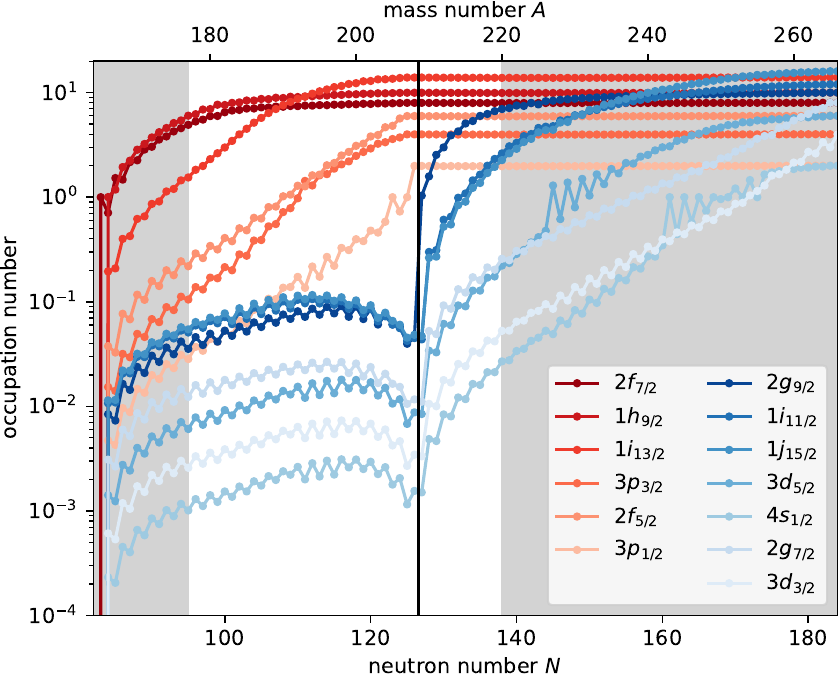}
\caption{
The same as in Figure~\ref{fig:n} but for all neutron fillings between $N=82$ and $N=184$ of the Pb isotopes. Red (blue) colors indicate the shells below (above) the magic gap. The gap itself is indicated by the black vertical line. The known Pb isotopes have neutron numbers within the white area.
}
\label{fig:alln}
\end{figure}

Here we examine the softening of the Fermi edge in Pb due to pairing by computing the occupation numbers $\avg{n_j} = \sum_{m} \avg{n_{jm}}$ of each $j$-shell. The results for half filling and 3/4 filling are shown in Figure~\ref{fig:n}. One observes that it is favorable not to fill the lowest two shells completely before putting a significant number of neutrons into the third $1i_{13/2}$ shell. Furthermore, it is favorable to place a small number of neutrons into much higher shells (even across the magic gap), in particular for high angular momenta.

Figure~\ref{fig:alln} shows all occupation numbers as a function of $N$. We find that an occupation of around $\avg{n_j}\sim 10^{-1}-10^{-2}$ is kept in the high-$j$ levels $2g_{9/2}$, $1i_{11/2}$, $1j_{15/2}$ before the $^{208}$Pb magic shell is filled. Apart from some even-odd oscillations, the rate of filling follows two different laws: The deep shells ($2f_{7/2}$, $1h_{9/2}$ below the magic gap and $2g_{9/2}$, $1i_{11/2}$, $1j_{15/2}$ above it) are filled very rapidly at first and then cross over to a slow saturation. The higher shells are filled exponentially (as seen by the straight lines in the logarithmic plot). Their occupancy is thus exponentially suppressed at first and then steeply increases.
This is most extreme in the case of the $3p_{1/2}$ level, which has the smallest pairing overlap with the other shells. The filling of this level abruptly changes by one between $N=125$ and shell closure at $N=126$. This `rapid filling' effect is responsible for the dip in the $\Delta^{\lr{3}}$ curve at $N=125$~\cite{Rowe_Wood_2010}.

\subsection{\label{app:model}Single-particle parameters}

To leading order, the fermions within a nucleus can be described by the single-particle Schr\"odinger equation with an effective potential. The Woods-Saxon potential~\cite{Woods_Saxon_1954} is a well-established choice for an undeformed nucleus with spherical symmetry. It is near-constant within the nucleus and has a diffuse boundary, which can be described by a formfactor which is identical to the Fermi function:
\begin{equation}
f_{R,a}\lr{r} = \frac{1}{\exp\lr{\frac{r-R}{a}}+1},
\end{equation}
where $a$ is the diffuseness of the surface and $R=R_0A^{1/3}$ is the nuclear radius. In the Seminole parametrization~\cite{Schwierz_etal_2007}, the effective single-particle Hamiltonian reads:
\begin{equation}\label{hamiltonian}
\begin{split}
H &= -\frac{\hbar^2}{2\mu}\Delta + U^c_R\lr{r} -U_0~g_{N,Z,\kappa}~f_{R,a}\lr{r} \\
&+\lambda\frac{ U_0 \hbar^2}{2\mu^2c^2r}\lr{\frac{\partial f_{R_{\text{SO}},a}\lr{r}}{\partial r}} \vec{l}\cdot\vec{s},
\end{split}
\end{equation}
with $\mu$ being the reduced mass, $c$ the speed of light, $U^c_R\lr{r}$ a Coulomb term which appears for protons only, $g_{N,Z,\kappa}$ an isospin-dependent prefactor ($N$: number of neutrons, $Z$: number of protons, $A=N+Z$, $\kappa$: free parameter), and $\vec{l}\cdot\vec{s}$ the spin-orbit coupling. The six parameters of the model are given by the potential depth $U_0=52.06~\text{MeV}$, isospin splitting $\kappa=0.639$, radius $R_0=1.260~\text{fm}$, diffuseness $a=0.662~\text{fm}$, dimensionless spin-orbit coupling $\lambda=24.1$, and spin-orbit radius $R_{0,\text{SO}}=1.16~\text{fm}$. The reader is referred to Ref.~\cite{Schwierz_etal_2007} for more details and a comparison to other parametrizations.

The eigenstates of the single-particle Schr\"odinger equation are given by
\begin{equation}
\psi_{nljm}\lr{r,\theta,\phi} = R_{nlj}\lr{r} i^{l} \mathcal{Y}_{jl}^m\lr{\theta,\phi},
\label{eq:Schroedinger_eigenstate}
\end{equation}
where $m=m_j$ is the projection of the total angular momentum $j$ on the z-axis, and $j$ is restricted by the relation $j=l\pm 1/2$ as a result of the action of the spin-orbit operator:
\begin{equation}
\vec{l}\cdot\vec{s}~\psi_{nljm} =
\left\{
\begin{array}{ll}
\frac{1}{2}l \psi_{nljm} & j=l+1/2, \\
-\frac{1}{2}\lr{l+1} \psi_{nljm} & j=l-1/2.
\end{array}
\right.
\end{equation}
$R_{nlj}\lr{r}$ is the solution of the radial Schr\"odinger equation, which we obtain numerically using a Chebyshev polynomial technique~\cite{Yucel_2015}. $\mathcal{Y}_{jl}^m\lr{\theta,\phi}$ are the spinor spherical harmonics that result from the coupling of the fermionic spin $s=1/2$ to the orbital momentum $l$:
\begin{equation}
\mathcal{Y}_{j=l\pm1/2}^{m}\lr{\theta,\phi} = 
\begin{pmatrix}
\alpha^{\uparrow}_{l,m,\pm} ~ Y_{l}^{m-1/2}\lr{\theta,\phi} \\
\alpha^{\downarrow}_{l,m,\pm} ~ Y_{l}^{m+1/2}\lr{\theta,\phi}
\end{pmatrix},
\end{equation}
where $\alpha^{\uparrow}_{l,m,\pm}=\pm\sqrt{\frac{l\pm m+1/2}{2l+1}}$ and $\alpha^{\downarrow}_{l,m,\pm}=\sqrt{\frac{l\mp m+1/2}{2l+1}}$ are the Clebsch-Gordan coefficients of this coupling, only the plus sign is possible for $l=0$; and $Y^{m}_l\lr{\theta,\phi}$ are the regular spherical harmonics (set to $Y^{m}_l=0$ for $m>l$).
The factor $i^{l}$ is a matter of convention, and is related to time-reversal~\cite{Huby_1954,Bohr_Mottelson_1998,Rowe_Wood_2010}. As is common practice, we subsume the indices $n$, $l$ and $j$ under one index $j\equiv\lr{n,l,j}$, which is understood to run over all split levels that are degenerate in $m$, resulting in a shorthand $\ket{jm}$ braket notation.

A normalized two-particle state in the shell $j$, where the two particles are coupled to a total angular momentum $J$ with a projection $M$ takes the form~\cite{Zelevinsky_Volya_2017}
\begin{equation}
\ket{jj;JM} = \frac{1}{\sqrt{2}}\sum_{mm'} C^{JM}_{jmjm'} a^{\dagger}_{jm} a^{\dagger}_{jm'} \ket{0},
\label{eq:app:jj;JM}
\end{equation}
where $a^{\dagger}_{jm}$ is a fermionic creation operator, $\ket{0}$ denotes the vacuum, $j$ is half-integer and $J$ is even integer. For the ground state, the dominant residual interaction term is given by singlet coupling $J=M=0$, which we can explicitly write out as~\cite{Zelevinsky_Volya_2003,Zelevinsky_Volya_2017,Liu_etal_2020}
\begin{equation}
\begin{split}
\ket{jj;00} &= \frac{1}{\sqrt{2}} \sum_{m=-j}^{j} \frac{\lr{-1}^{j-m}}{\sqrt{2j+1}} a^{\dagger}_{j,+m}a^{\dagger}_{j,-m} \ket{0}\\
            &= \frac{1}{\sqrt{j+\frac{1}{2}}} \sum_{m>0} \lr{-1}^{j-m} a^{\dagger}_{j,+m}a^{\dagger}_{j,-m} \ket{0}.
\end{split}
\label{eq:singlet}
\end{equation}
The second-quantized Hamiltonian that describes hoppings of singlet pairs between all $j$-shells with the strengths $G_{jj'}$ takes the following form:
\begin{equation}
\begin{split}
H = &\sum_{j,m>0}\epsilon_{j} n_{jm} - \sum_{jj'} G_{jj'} \sum_{m>0} \lr{-1}^{j-m} a^{\dagger}_{jm}a^{\dagger}_{j,-m}\\
& \times \sum_{m'>0} \lr{-1}^{j'-m'} a_{j',-m'} a_{j'm'}.
\label{eq:app:H}
\end{split}
\end{equation}
The couplings $G_{jj'}$ are {\it a priori} unknown. One should note that in order to properly describe excited states, it is necessary to add further quartic terms to the Hamiltonian.


%

\end{document}